\documentclass[onecolumn]{aastex631}

\usepackage{amsmath} 
\usepackage{subfigure}
\usepackage{float}

\newcommand{\CASTRO}{\texttt{CASTRO}}
\newcommand{\KEPLER}{\texttt{KEPLER}}
\newcommand{\STELLA}{\texttt{STELLA}}
\newcommand{\ATHENA}{\texttt{ATHENA}}
\newcommand{\HERACLES}{\texttt{HERACLES}}
\newcommand{\FRONT}{\texttt{FRONT}}
\newcommand{\RAMSES}{\texttt{RAMSES}}
\newcommand{\RAGE}{\texttt{RAGE}}
\newcommand{\SPECTRUM}{\texttt{SPECTRUM}}
\newcommand{\Ni}{{\ensuremath{^{56}\mathrm{Ni}}}}
\newcommand{\Fe}{{\ensuremath{^{56}\mathrm{Fe}}}}

\newcommand{\He}{{\ensuremath{^{4} \mathrm{He}}}}
\newcommand{\Hy}{{\ensuremath{^{1} \mathrm{H}}} }
\newcommand{\Ox}{{\ensuremath{^{16}\mathrm{O}}}}

\newcommand{\Ca}{{\ensuremath{^{40}\mathrm{Ca}}}}

\newcommand{\Si}{{\ensuremath{^{28}\mathrm{Si}}}}
\newcommand{\Ti}{{\ensuremath{^{44}\mathrm{Ti}}}}
\newcommand{\Ne}{{\ensuremath{^{20}\mathrm{Ne}}}}
\newcommand{\Mg}{{\ensuremath{^{24}\mathrm{Mg}}}}
\newcommand{\Cx}{{\ensuremath{^{12}\mathrm{C}}}}

\newcommand{\erg}{{\ensuremath{\mathrm{erg}}}}

\newcommand{\K}{\ensuremath{\mathrm{K}}}
\newcommand{\gcc}{\ensuremath{\mathrm{g}\,\mathrm{cm}^{-3}}}
\newcommand{\cms}{\ensuremath{\mathrm{cm}\,\mathrm{s}^{-1}}}

\newcommand{\Ms}{{\ensuremath{\mathrm{M}_{\odot} }}}

\newcommand{\rev}[1]{\textcolor{black}{ #1}}

\newcommand{\Ep}[1]{{\ensuremath{10^{#1}}}}
\newcommand{\E}[1]{{\ensuremath{\powersep\Ep{#1}}}}
\newcommand{\B}[1]{{\ensuremath{\left({#1}\right)}}}
\newcommand{\powersep}{{\ensuremath{\times}}}

\makeatletter

\newcommand{\Rmnum}[1]{\expandafter\@slowromancap\romannumeral #1@}

\shorttitle{Pulsational Pair-Instability Supernovae}
\shortauthors{Chen et al.}

\graphicspath{{./}{figures/}}

\begin{document}

\title{Multidimensional Radiation Hydrodynamics Simulations of Pulsational Pair-Instability Supernovae}

\correspondingauthor{Ke-Jung Chen}
\email{kjchen@asiaa.sinica.edu.tw}

\author{Ke-Jung Chen}
\affiliation{Institute of Astronomy and Astrophysics, Academia Sinica, Taipei 10617, Taiwan R.O.C.}

\author{Daniel J. Whalen}
\affiliation{Institute of Cosmology and Gravitation, University of Portsmouth, Portsmouth PO1 3FX, UK}

\author{S. E. Woosley}
\affiliation{Department of Astronomy and Astrophysics, University of California at Santa Cruz, Santa Cruz, CA 95060, USA}
	
\author{Weiqun Zhang}
\affiliation{Center for Computational Sciences and Engineering, Lawrence Berkeley National Lab, Berkeley, CA 94720, USA}

\begin{abstract}

Stars with masses of 80 - 130 \Ms\ can encounter the pulsational pair-instability at the end of their lives, which triggers consecutive episodes of explosive burning that eject multiple massive shells. Collisions between these shells produce bright transients known as pulsational pair-instability supernovae (PPI SNe) that may explain some extreme supernovae. In this paper, we present the first 2D and 3D radiation hydrodynamics simulations of PPI SNe with the \CASTRO\ code. Radiative cooling causes the collided shells to evolve into thin, dense structures with hot spots that can enhance the peak luminosity of the SN by factors of 2 - 3. The light curve peaks at $1.9 - 2.1 \times 10^{43}$ erg s$^{-1}$ for 50 days and then plateaus at $2 - 3 \times 10^{42}$ erg s$^{-1}$ for 200 days, depending on viewing angle.  \rev{The presence of \Cx\ and \Ox\ and absence of \Si\ and \Fe\ in its spectra can uniquely identify this transient as a PPI SN in follow-up observations. Our models suggest that multidimensional radiation hydrodynamics is required to model the evolution and light curves of all shell-collision SNe such as Type IIne, not just PPI SNe.}

\end{abstract}

 
\section{Introduction}

Stars with masses $\gtrsim$ 80 \Ms\ can build up helium cores exceeding 35 \Ms\ that encounter the pair-creation instability at the end of their lives. Pair-creation converts pressure-supporting photons into electron-positron pairs and causes the core to contract during central oxygen burning. Core contraction raises the central temperature and ignites explosive oxygen burning that does not unbind the star but causes the core to pulsate with a period of a few hundred seconds and produce many weak shocks \citep{w17,r22}. As the helium core grows in mass, the pulses become less frequent but more energetic. They can trigger a few giant eruptions and produce supernova (SN)-like transients at helium core masses above 45 \Ms. The first strong pulse has energies of  $10^{49}-10^{50}$ \erg\ that easily eject the hydrogen envelope, whose gravitational binding energy is $10^{43}-10^{44}$ \erg, and produce a faint Type IIP SN. However, when subsequent eruptions collide with the first, they may produce a much brighter Type IIn SN. If no hydrogen envelope remains during the explosions, collisions between helium shells ejected by the pulsations can make a luminous Type I SN. For helium core masses of $45 - 55$ \Ms, the interval between pulses becomes several years, and the shells collide at radii of $10^{15}-10^{16}$ cm if they have velocities of  $\sim 10^{8}\, \cms$.  In these circumstances, much of the collision energy is dissipated as optical emission known as a pulsational pair-instability supernova \citep[PPI SN;][]{brk67,wbh07,w17}.  

\citet{wbh07} modeled the PPI SN of a 110 \Ms\ solar-metallicity star and found that the collision of a 10$^{51}$ erg eruption can yield a superluminous SN \citep[SLSN;][]{ins16,tak18,chen21} such as SN 2006gy.  Followup studies by \citet{w17} and \citet{lnb19} found a broad range of outcomes for PPI explosions, from multiple faint SNe to a single SLSN \citep{wbh07,wet13d,des15,ml15,jer16,w17}. A PPI SN has also been used to explain the light curve (LC) of SN IPTF14hls \citep{arc17} which has extended multiple peaks that could be due to collisions between shells \citep{w18}.  

However, in 1D simulations much of the luminosity originates from a thin, dense shell that forms during the collision that would be prone to fragmentation and clumping by hydrodynamical (HD) instabilities in 2D and 3D.  2D HD simulations of a PPI SN by \citet{chen14a} have shown that Rayleigh-Taylor (RT) instabilities drive mixing between the colliding shells and break apart the dense shell. However, this 2D study could not evaluate the effect of the RT instabilities on the LC and spectra of the collision because it did not include radiation hydrodynamics (RHD), which is required to properly model radiation flow through the ejecta and to determine how gas heating and radiative cooling affect fragmentation. Realistic LC calculations for PPI SNe must therefore be performed in at least 2D on a mesh that is fine enough to resolve the thin radiating regions of the collisions.

We have performed multidimensional RHD simulations of a PPI SN with the \CASTRO\ code to study how radiation flows alter the structure of the dense shell and how instabilities affect the LC of the collision. We compare our results with previous simulations with only hydrodynamics. We describe the setup for our \CASTRO\ simulations in Section 2 and present our 1D, 2D, and 3D models in Sections 3, 4, and 5, respectively. We discuss the observational
signatures of these events, which include the effects of multidimensional RHD mixing in Section 6 and conclude in Section 7.

\section{Numerical Method}

\subsection{\KEPLER{}}

We initialize our \CASTRO\ simulations with the 110 \Ms\ solar-metallicity PPI SN that was originally modeled in \KEPLER\ \citep{kep1,kep2} by \citet{wbh07}) and further studied by \citet{wet13d} and \citet{chen14a}.  Mass loss is suppressed to a fraction of the typical value for a solar-metallicity star, 50\% and 10\% in the main and post-main sequences, respectively. The star has a mass of 74.6 \Ms\ and a 49.9 \Ms\ helium core when it encounters the PI as a red supergiant, with a radius of $1.1 \times 10^{14}$ cm and a luminosity of 9.2 $\times$ 10$^{39}$ erg s$^{-1}$. Pair creation causes the core to contract and increase in temperature to $3.04 \times 10^9$ K, well above the stable oxygen-burning temperature of $\sim 2.0 \times 10^9$ K in massive stars. This rise in temperature triggers explosive burning that consumes 1.55 \Ms\ of \Cx\ and 1.49 \Ms\ of \Ox\ and releases 1.4 $\times$ 10$^{51}$ erg of energy. 90\% of this energy expands the star and the rest ejects the outer layers of the core and surrounding envelope: 17.2 \Ms\ of \He\ and 7.3 \Ms\ of \Hy. The ejected shell has peak velocities of $\sim$ 10$^8$ cm s$^{-1}$. As shown in Figure 2 of \citet{wbh07}, the expulsion of the envelope produces a faint SN with a brief breakout luminosity of 5 $\times$ 10$^{42}$ erg s$^{-1}$ followed by a 7.9 $\times$ 10$^{41}$ erg s$^{-1}$ plateau powered primarily by \He\ and \Hy\ recombination. We call this first pulse P1. What remains after P1 is a 50.7 \Ms\ star that is slightly more massive than the original helium core of 49.9 \Ms.

After 6.8 years, the helium core again contracts and encounters the pair-instability twice in a rapid succession, ejecting two strong pulses P2 and P3. The total mass and kinetic energy of P2 and P3 are 5.1 \Ms\ and $6.0 \times 10^{50}$ erg.  P3 later overtakes P2, and their merged shell (P2+P3) eventually collides with P1. Nine years later, the core contracts again and enters a stable silicon burning phase that forges a massive iron core that collapses directly to a black hole without a SN or gamma-ray burst \citep[GRB;][]{wet08c,met12a,mes13a}. We map the 1D \KEPLER\ profiles of this star onto the \CASTRO\ grids when P3 is launched and the forward shock of P1 is at $r \sim 5 \times 10^{16}$ cm. The shell collision is evolved until the forward shock of P3 reaches $r \sim 10^{16}$ cm, after most of the thermal energy from its collision with P2 has been dissipated by radiative cooling \citep{wbh07}.

\rev{Although our 110 \Ms\ star is not primordial in composition, its energetics are representative of PPI SNe whose intervals between ejections produce bright transients in optical and UV in the rest frame, which are most likely to be detected at high redshifts today.  Its evolution also approximates that of low-metallicity stars because mass loss was suppressed over its lifetime.  The large grid of progenitor masses, metallicities, and prescriptions for mass loss studied by \citet{w17} confirmed that only stars with final helium core masses above 45 \Ms\ produce energetic eruptions and SN-like transients.  We therefore consider our 110 \Ms\ PPI SN as a fiducial case.}  

\subsection{\CASTRO{}}

\CASTRO\ is a massively-parallel multidimensional AMR RHD code for astrophysical simulations \citep{alm10}.  It uses an unsplit piecewise-parabolic method (PPM) HD scheme \citep{wc84} to avoid spurious noise due to dimensional splitting. We use the ideal gas law for the equation of state in our \CASTRO\ simulations, which is suitable for the low-density gas ($\rho < 10^{-10} \gcc$) in colliding shells. 1D \KEPLER\ profiles of velocities, temperatures, densities, and elemental mass fractions are mapped onto the AMR grid hierarchy in \CASTRO\ with a conservative algorithm that strictly preserves these quantities on the new grids \citep{chen11,chen13}.  The gravity solver uses the monopole approximation by constructing a 1D spherical profile of the gas density and using it to calculate gravitational forces in all AMR grids.  This approximation is reasonable because there are no serious departures from spherical symmetry in PPI SN ejecta.  We only track \Hy, \He, \Cx, and \Ox\ because other heavier elements such as \Ne, \Mg, \Si, \Ca, \Ti, and \Ni\ are mostly absent in PPI SNe.   

\subsection{RHD}

The \CASTRO\ RHD module solves two-temperature, multi-group flux-limited diffusion (MGFLD) in which gas temperatures can differ from radiation temperatures.  It adopts a mixed frame solution to the RHD equations assuming local thermodynamic equilibrium \citep[LTE;][]{zhang11,zhang13} and a second-order explicit Godunov method for the hyperbolic part of the system and a first-order backward Euler method for the parabolic part. The mixed-frame approach in \CASTRO\ is similar to that in {\tt Orion} \citep{krum07} and has the advantage of being strictly conservative in energy. The RHD module in \CASTRO\ has been well tested and applied to a number of astrophysical simulations such as neutrino-driven core-collapse SNe \citep{dol15} and shock breakout in SNe \citep{lov17}.  

Here, we use the gray approximation based on a frequency-integrated formulation of the RHD equations.  For simplicity, we use the frequency and temperature independent electron scattering opacity, $\kappa$ and consider four cases: $\kappa$ =  0.1, 0.2, 0.3, and 0.4 ${\rm cm}^2\,{\rm g}^{-1}$ to sample different ionized fractions in the gas. The LCs are calculated by tallying radiation flux at the photosphere, where the photons become free-streaming,
\begin{equation}
L = 4\pi r_{\rm p}^2 F_{\rm rad},
\end{equation}
where $L$ is the bolometric luminosity in erg s$^{-1}$ \rev{and $F$ is the radiation flux in erg cm$^{-2}$ s$^{-1}$ at $r_{\rm p}$, the radius of the photosphere, which is beyond the outer edge of the dense shell.} 

Our 1D \CASTRO\ simulations are performed on a spherical coordinate grid with reflecting and outflow boundary conditions at the inner and outer boundaries at $r =$ 0 and 2 $\times$ 10$^{16}$ cm, respectively.  To test different resolutions, we use uniform grids with 1024, 4096, and 8192 zones. In 2D runs, we simulate one quadrant of the star on a cylindrical coordinate mesh in $r$ and $z$ and apply outflow and reflecting boundary conditions to the upper and lower boundaries in $r$ and $z$, respectively. The base grid in our two 2D runs has 256$^2$ and 512$^2$ zones with three levels of factor-two refinement (2$^3$) that yield an effective maximum resolution of (256 $\times$ 2$^3$)$^2$ $=$ 2048$^2$ and (512 $\times$ 2$^3$)$^2$ $=$ 4096$^2$ on a square that is $2 \times 10^{16}$ cm on a side. Our 3D \CASTRO\ runs are performed on a Cartesian coordinate grid. We center the full star at $x=0$, $y=0$, and $z=0$ and apply outflow boundary conditions to all outer boundaries. The root grid has 512$^3$ zones with up to two levels of refinement (2$^2$) for an effective resolution of (512 $\times$ 2$^2$)$^3$ $=$ 2048$^3$ in a cube that is $2 \times 10^{16}$ cm on a side. All AMR grids were refined on gradients in density and velocity, and all the simulations were evolved for 350 days after the ejection of P3.

\section{1D PPI SN Evolution}

% Figure 1 

\begin{figure*}
\begin{center}
\includegraphics[width=\textwidth]{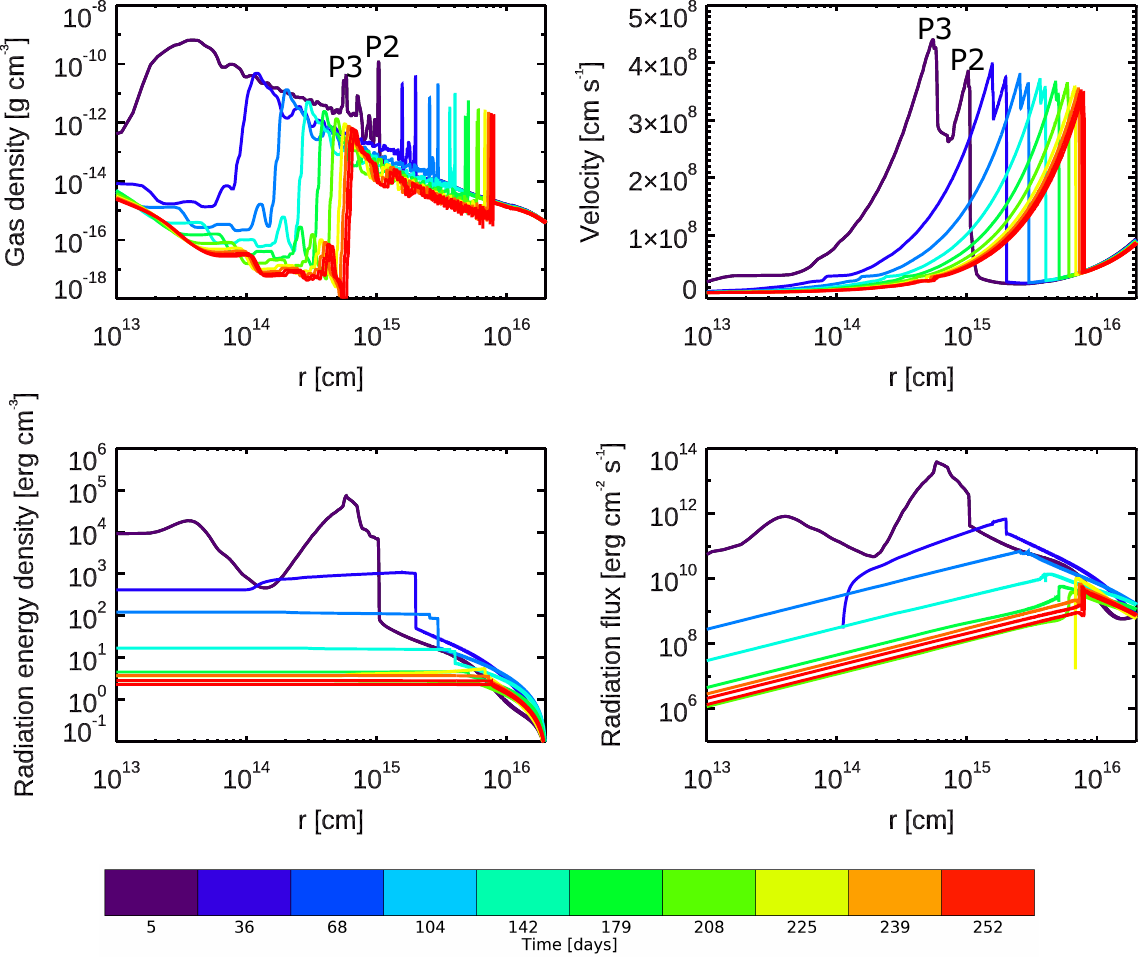} 
\caption{Gas densities, velocities, radiation energy densities, and radiation fluxes for the $\kappa = 0.1$, 8192-zone model from 0 - 350 days after the ejection of P3. P2 and P3 mark the positions of the shocks in the corresponding shells. At $t = 0$, the gas at $r > 10^{15}$ cm is the tail of P1. The collision between P3 and P2 can be seen in the merger of the two velocity peaks over time. Collisional heating produces a large radiation flux at the shock front P2.
\label{fig:1d_evol}}
\end{center}
\end{figure*}

% Figure 2 

\begin{figure}
\begin{center}
\includegraphics[width=.6\columnwidth]{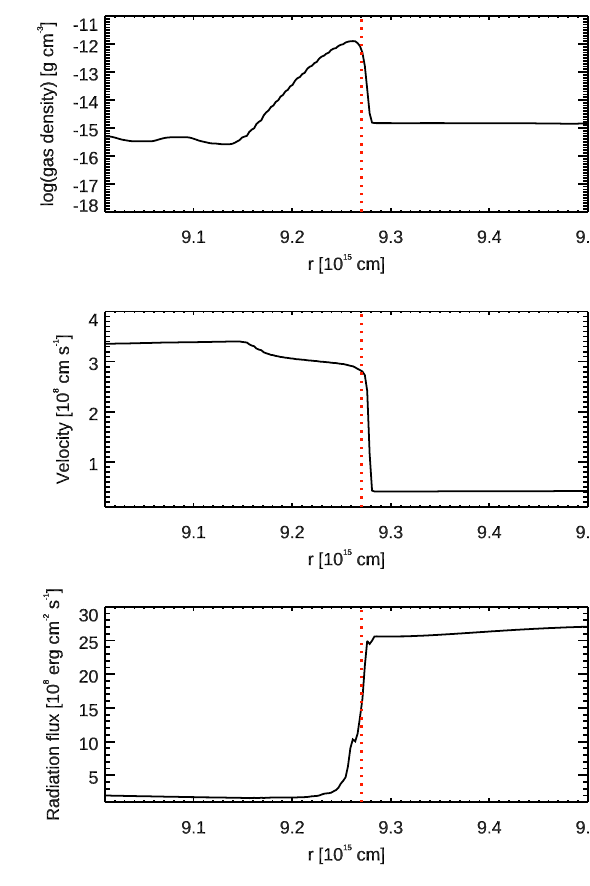} 
\caption{Zoom-in of the density spike in the 1D $\kappa = 0.1$, 8192-zone run at 300 days. The red dashed line marks the position of the shockin P2+P3.  The spike is $\sim 1.4 \times 10^{14}$ cm across and is resolved by about 50 zones. The density of the spike peaks at $\sim 10^{-12} \gcc$, about 1,000 times higher than its surroundings. The spike also exhibits  abrupt changes in velocity and radiation flux.
\label{fig:spike_zoom}}
\end{center}
\end{figure}

% Figure 3 

\begin{figure}
\begin{center}
\begin{tabular}{cc}
\includegraphics[width=.5\columnwidth]{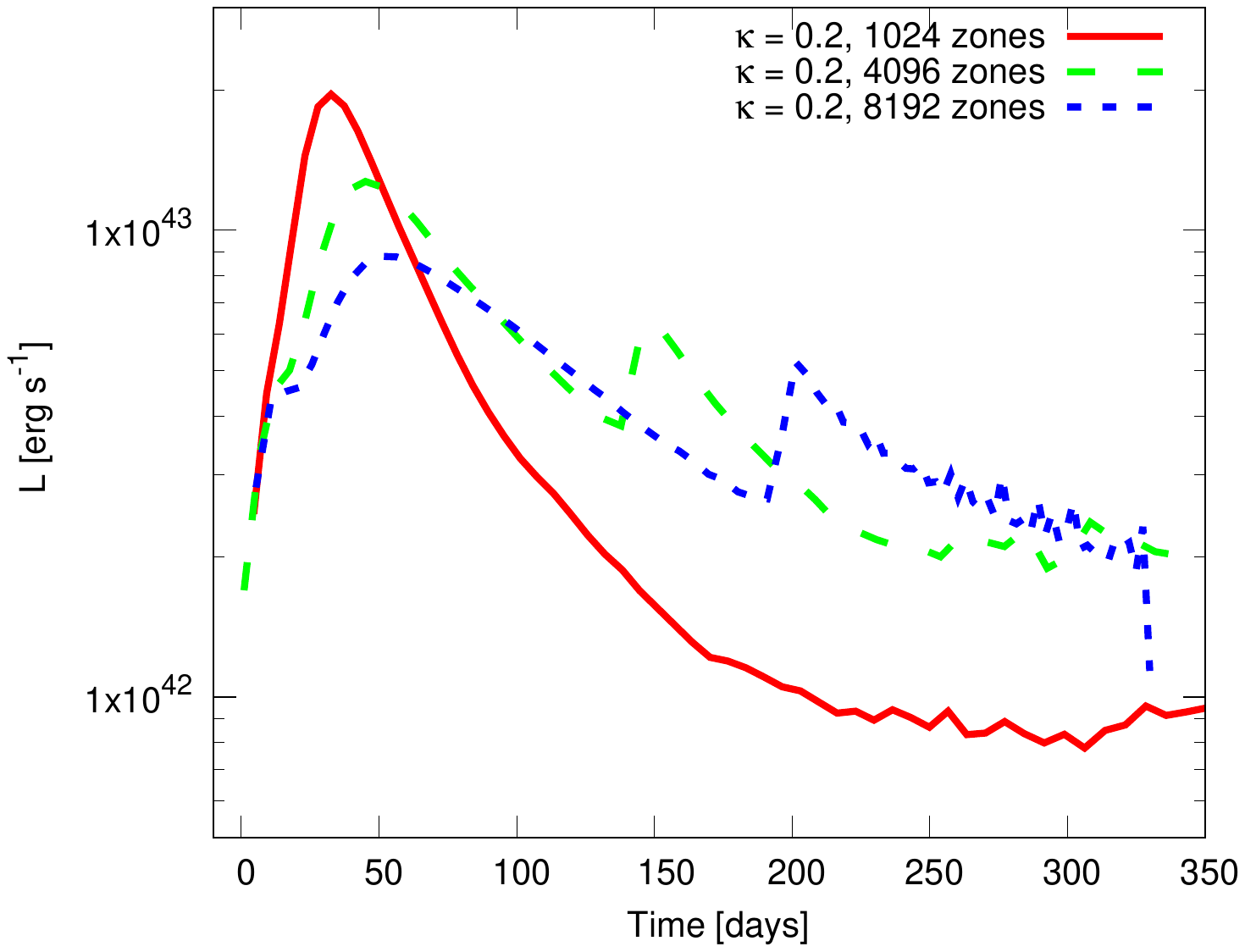}  &
\includegraphics[width=.5\columnwidth]{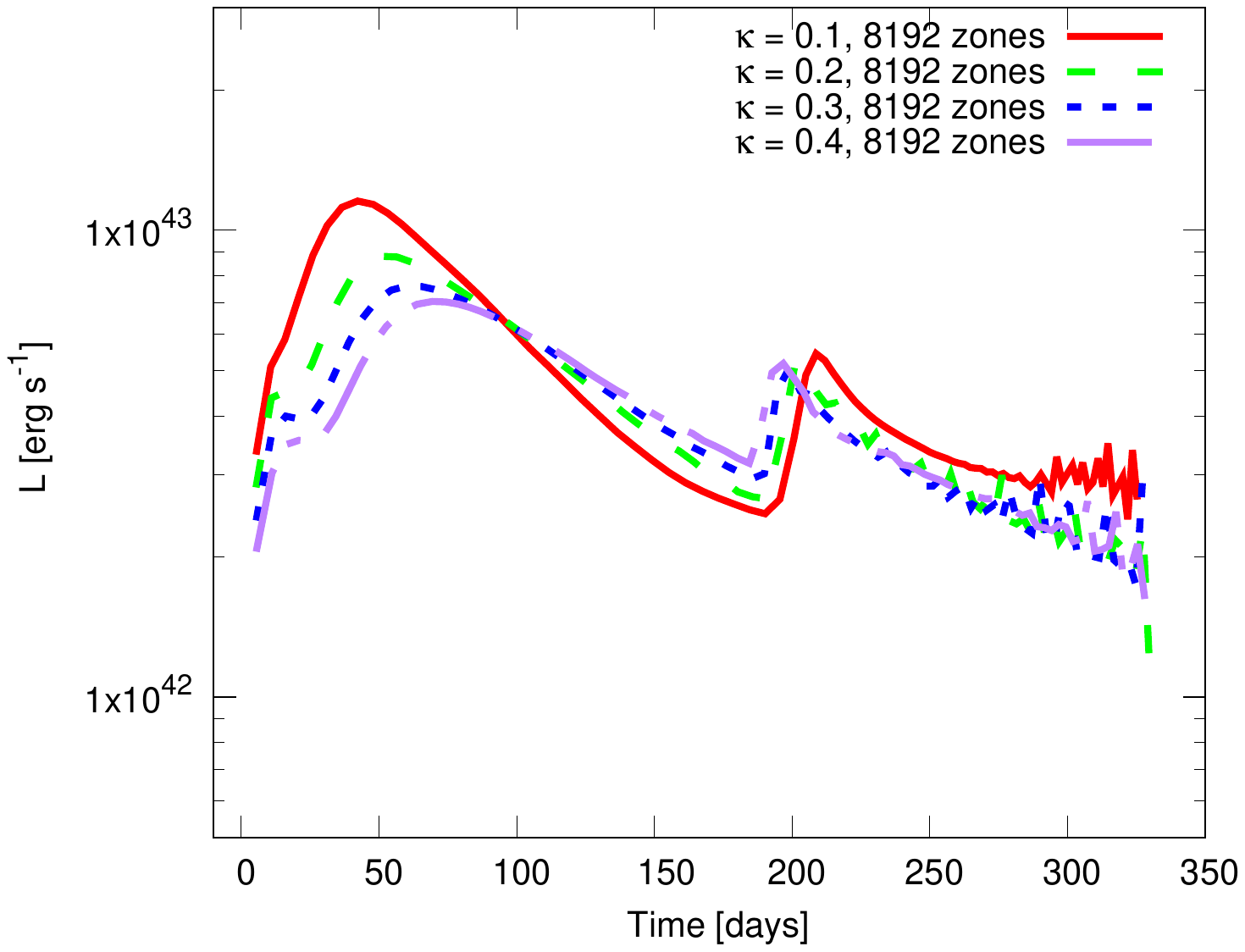}  
\end{tabular}
\caption{LCs for the 1D runs.  \emph{Left}: 1024, 4096, and 8192 zones with $\kappa = 0.2$.  The second peak is smeared out in the 1024 zone run because the thin shells are underresolved. \emph{Right}:  LCs for the 8192 zone run for $\kappa =$ $0.1 - 0.4$.  
\label{fig:1d_lc}}
\end{center}
\end{figure}

We show density, velocity, radiation energy density, and radiation flux profiles for the 1D run in Figure~\ref{fig:1d_evol}. At the beginning of the simulation (the launch of P3), the shock in P2 is at $r = 7.9 \times 10^{14}$ cm and has a radial velocity of $\sim 3.9 \times 10^8$ cm s$^{-1}$.  The shock in P2 is at $r = 3.2 \times 10^{14}$ cm and has a radial velocity of $\sim 4.8 \times 10^8$ cm s$^{-1}$. At this time, the forward shock of P1 is at $r = 5 \times 10^{16}$ cm and is not visible here.  As P2 plows through the tail of P1, it decelerates and shocks the gas, piling up the hot, thin, dense shell that is visible as the large density spike with an overdensity $<\delta \rho / \rho> \sim 1,000$ at $r = 7.9 \times 10^{14}$ cm with several smaller spikes between P2 and P3.  When P3 begins to catch up to P2 and plow up its tail, it forms another large density spike, albeit with a smaller $<\delta \rho / \rho>$ of $\sim 300$.  The collision of the ejected shells converts part of their eruption energy into light.  As shown in the lower two panels of Figure~\ref{fig:1d_evol}, the loci of maximum radiation flux coincide with the shocks in P2 and P3.  

In the \citet{chen14a} HD simulation, P3 overtakes P2 in 50 days, merging with it at $r \sim$ 2.3 $\times$ 10$^{15}$ cm before they both plow up the tail of P1.  However, in 1D RHD simulations, radiative cooling dissipates the energy of the eruption and decelerates the ejecta, delaying the merger of P2 and P3 until $r \sim 5.1 \times 10^{15}$ cm at $\sim 130$ days. This result is consistent with \citet{wet13d}, who studied the same explosion in 1D with the \RAGE\ and \SPECTRUM\ codes \citep{rage,fet12} with the OPLIB atomic opacities \citep{oplib}.  They found that P2 and P3 merge at 150 days at $r \sim 6 \times 10^{15}$ cm.  \rev{The P2+P3 shell creates a large density spike that was moderately resolved in 1D Lagrangian RHD simulations with \STELLA\ \citep{stella,wbh07,moriya12} -- see, e.g., Figure 7 in \citet{moriya12} -- but poorly resolved in \KEPLER\ \citep{kep2,w17}}.  We show a closeup of the spike for the $\kappa = 0.1$, 8192-zone run in Figure~\ref{fig:spike_zoom}.  It has a width of $\sim 1.4 \times 10^{14}$ cm with a jump in density from $\sim 8 \times 10^{-16}$ \gcc\ to $\sim 1.1 \times 10^{-12}$ \gcc. 
 
We show LCs for the 1D models for $\kappa = 0.2$ and 1024, 4096, and 8192 zones in the upper panel of Figure~\ref{fig:1d_lc}. The LC of the 1024-zone run has the highest peak luminosity and shortest duration.  A second peak appears in the 4096 and 8192-zone LCs, and it is earlier in the 4096-zone run. Grid resolution affects the LCs because it determines how well the shells are resolved before and after the collisions.  The peak luminosities are $0.7-1.3 \times 10^{43}$ erg s$^{-1}$ and are powered by the collision of P3 with P2.  The second peak appears at 200 days when the thin shells in P3 and P2 finally collide and merge, as shown in the 4096 and 8192-zone runs.  As shown in the 8192-zone run in the bottom panel of Figure~\ref{fig:1d_lc}, as the opacity increases so do the radiation diffusion times, and peak luminosities become broader and dimmer.

The peak luminosities of our LCs are somewhat lower than in \citet{wbh07} and \citet{wet13d}: $3.6 \times 10^{43}$ erg s$^{-1}$ and $6.9 \times 10^{43}$ erg s$^{-1}$, respectively.  \rev{These differences are likely due to opacity, physics and resolution. In \RAGE, \citet{wet13d} used OPLIB, 2T radiation transport, and over 100 times the resolution used here. They showed that the temperature of the gas in the thin shell was a few 10$^3$ K to 10$^5$ K and was never fully ionized, so atoms would definitely have contributed opacity to the radiation flow, not just free electrons. Our use of FLD was less of a factor in the differences between peak luminosities because \citet{wet13d} also used grey FLD and our peak brightnesses were a factor of a few lower than theirs. We note that the higher-order M1 scheme has now been applied to calculations of superluminous SN LCs in other radiation hydrodynamics codes such as \RAMSES\ \citep{RAMESE,ramesert}, \ATHENA\ \citep{ATHENA,vet}, \HERACLES\ \citep{HERACLES}, and \FRONT\ \citep{FRONT,frontrt} and may better model photon transport in regions of the flow that are intermediate to optically thick and thin regimes.}

Our LCs are smoother than those in \citet{wbh07} and \citet{wet13d}, which exhibit the classic ripples due to the radiative instability described by \citet{chev82} and \citet{imam84}.  As noted above, a reverse shock forms as the forward shock in P3+P2 plows up gas in the tail of P1.  The reverse shock initially backsteps into the flow in the rest frame of the forward shock.  However, if the postshock gas can radiatively cool, the reverse shock loses pressure support and catches back up to the forward shock.  The cycle repeats as the forward shock plows up more material and the reverse shock again recedes into the flow, imprinting oscillatory fluctuations in the LCs.  These fluctuations do not appear in 2D or 3D RHD simulations because, as we show in the next section, hydrodynamical instabilities form in the region between the colliding shells, disrupt the radiative instabilities, and smoothen the ripples in the LCs.

Although we do not calculate spectra here,  \citet{wet13d} found that the shock is hottest, $\sim$ 5 eV or 55,000 K, at $\sim$ 70 days and that its spectrum cuts off at about 500 \AA.  P2 therefore radiates strongly in the UV and the collision looks like a Type IIn SN. After P2 and P3 collide, when part of their kinetic energy has been dissipated the shock cools and softens its spectrum to optical wavelengths. The PPI SN here is similar to the Type IIn SNe in \citet{wet12e}, which exhibit early bright UV emission when the shock has large radii, $\sim$ 10$^{15}$ cm. This radius is comparable to those at which ejecta from the explosion crash into circumstellar material (CSM) in Type IIn events \citep{ml15, mor18}. With similar shock temperatures and radii upon collision, the initial total luminosities for PPI SNe and Type IIn are comparable, as shown in Figure 2 in \citet{wet12e}.  However, Type IIn SNe are powered by the collision between the ejecta and a circumstellar shell due to stellar winds before the massive star dies. CSM interactions are more abrupt than PPI eruptions because the densities of CSM have distinct boundaries. The duration of Type IIn SNe LCs is also shorter because the CSM mass is less than the shell masses in PPI SNe.

\section{2D PPI SN Evolution}

% Figure 4 

\begin{figure*}
\begin{center}
\includegraphics[width=.8\textwidth]{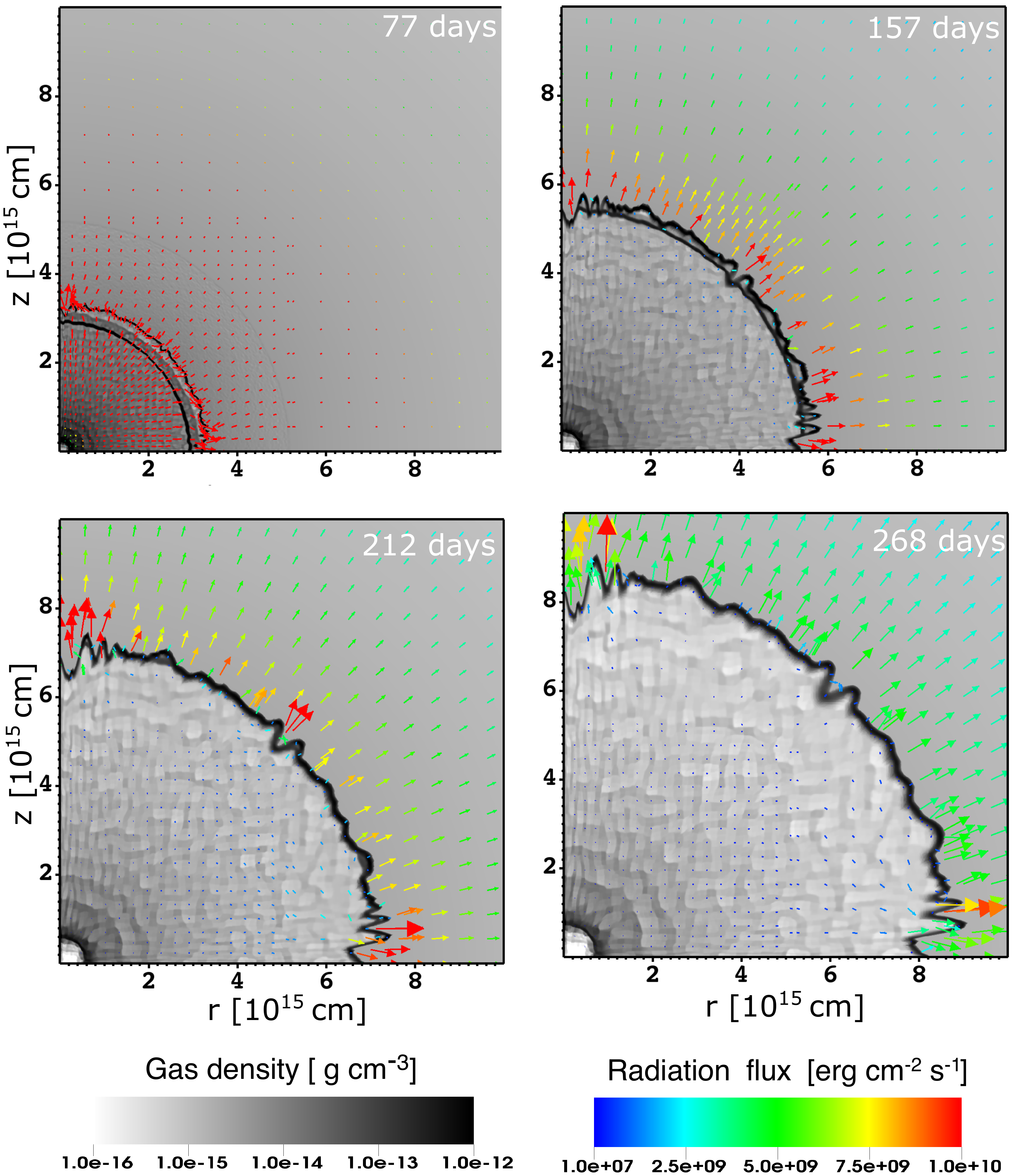} 
\caption{Evolution of the gas density and radiation flux in 2D for the $\kappa = 0.2$, 512$^2$ AMR run at 77, 157, 212, and 268 days, respectively. Initially, two dense shells form in P2 and P3 and later merge. The off-axis corrugations in the merged shell are due to radiative cooling.  \label{fig:2d_den}}
\end{center}
\end{figure*}

% Figure 5 

\begin{figure*}
\begin{center}
\includegraphics[width=\textwidth]{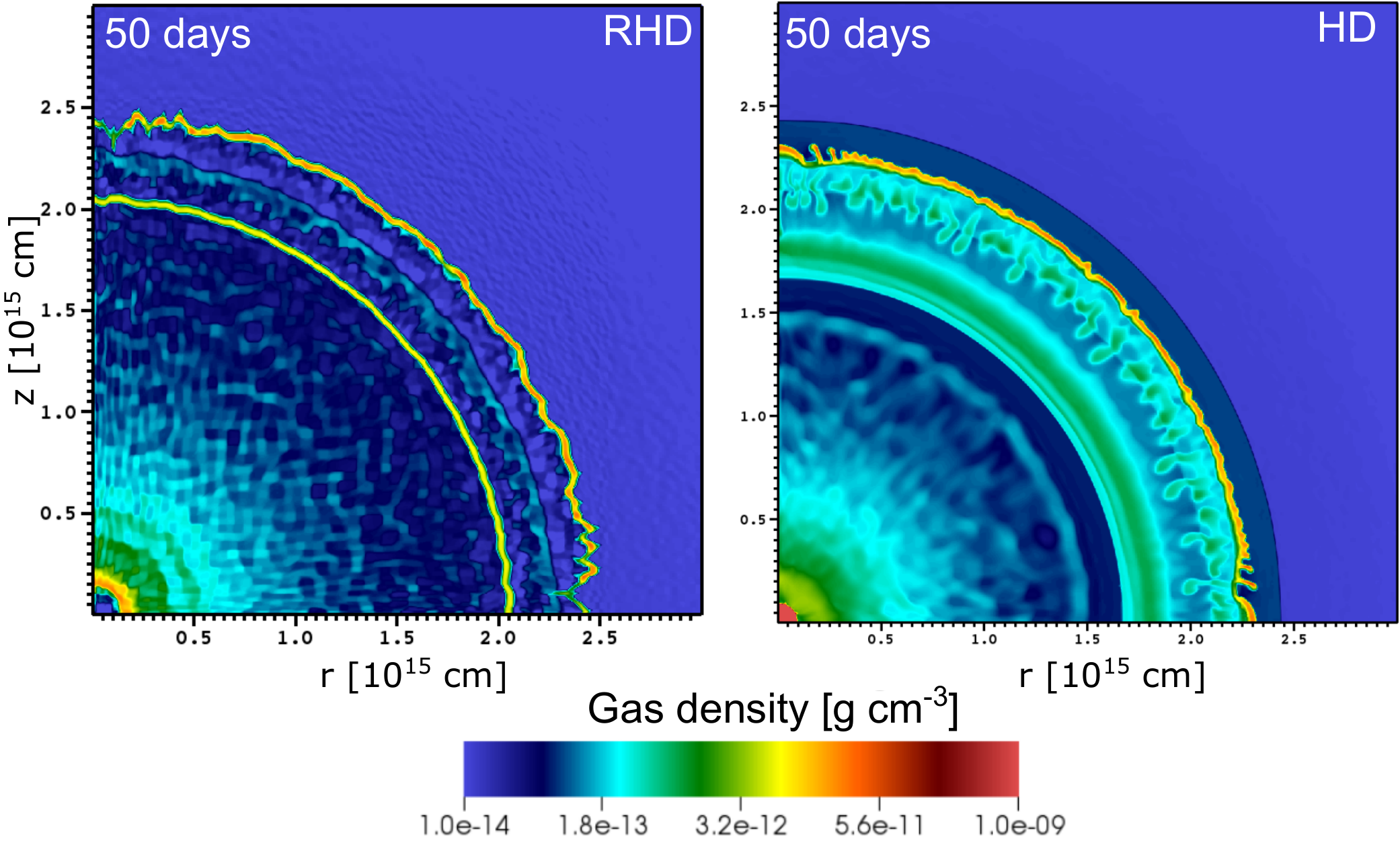} 
\includegraphics[width=\textwidth]{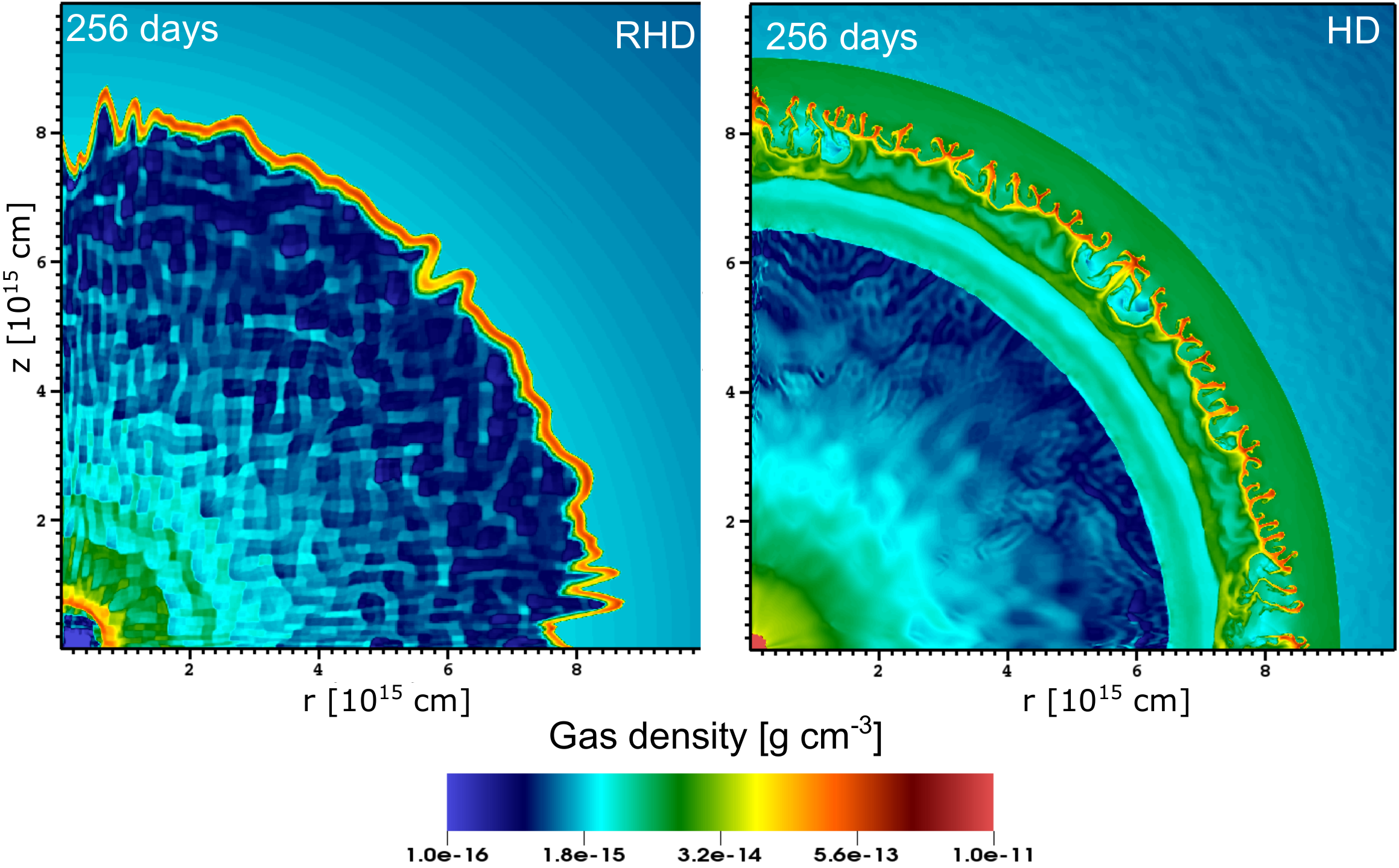} 
\caption{2D density images for the $\kappa = 0.2$, 512$^2$ AMR run (left) and the hydro-only HD model (right) from \citet{chen14a} at 50 and 256 days. The 2D RHD and HD simulations have resolutions of $4.88 \times 10^{12}$ cm and $1.22 \times 10^{12}$ cm, respectively. Mixing during the collision is stronger in HD than in RHD.
\label{fig:comp2d}}
\end{center}
\end{figure*}

% Figure 6 

\begin{figure*}
\begin{center}
\includegraphics[width=\textwidth]{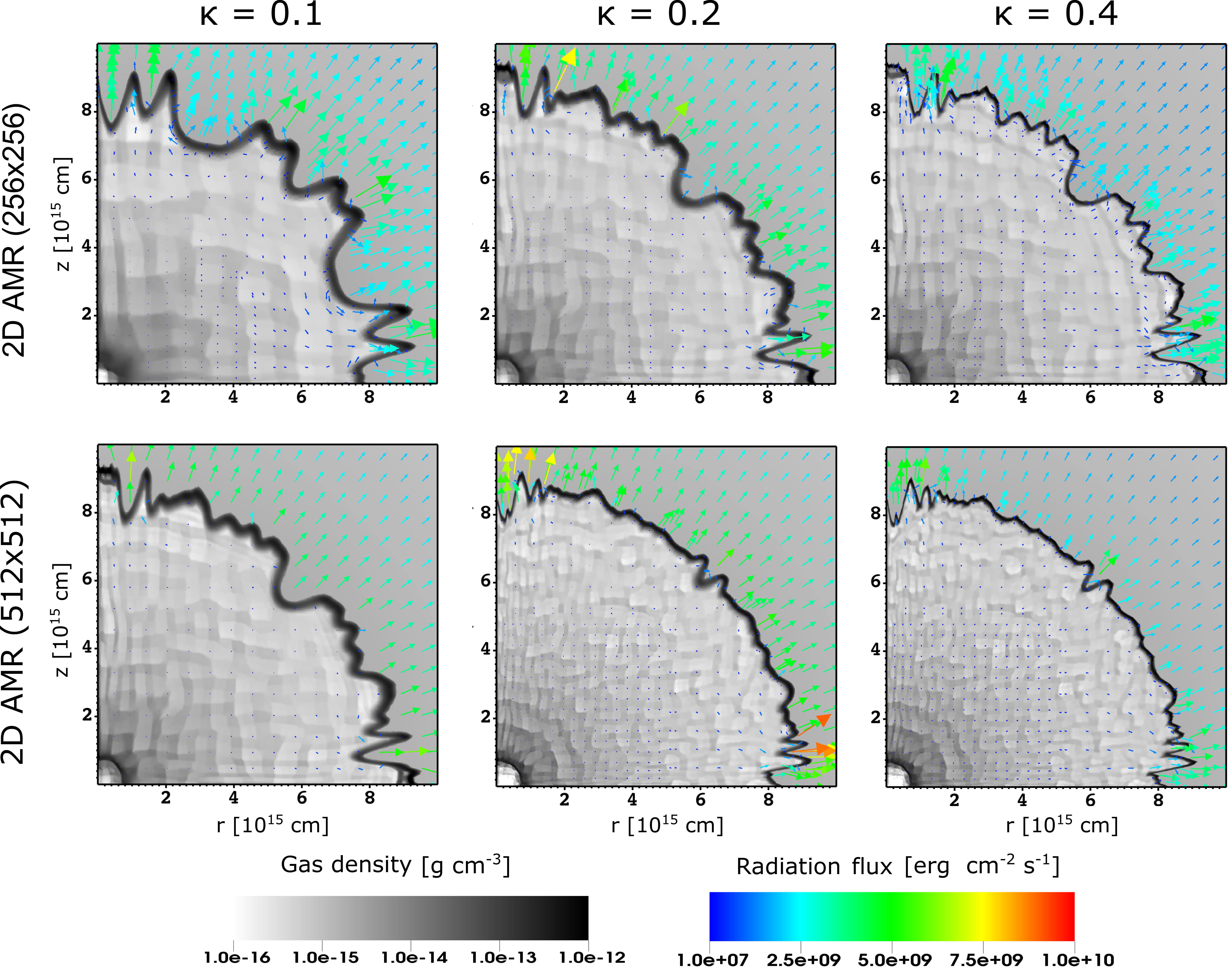} 
\caption{Density structure of the 2D $\kappa =$ 0.1, 0.2, and 0.4 runs at two different resolutions at 270 days.  The colored arrows mark the direction of radiation flux. The thickness and deformation of the shell decreases as the opacity increases.    
\label{fig:2d_comp1}}
\end{center}
\end{figure*}

% Figure 7 

\begin{figure}
\begin{center}
\includegraphics[width=1.\columnwidth]{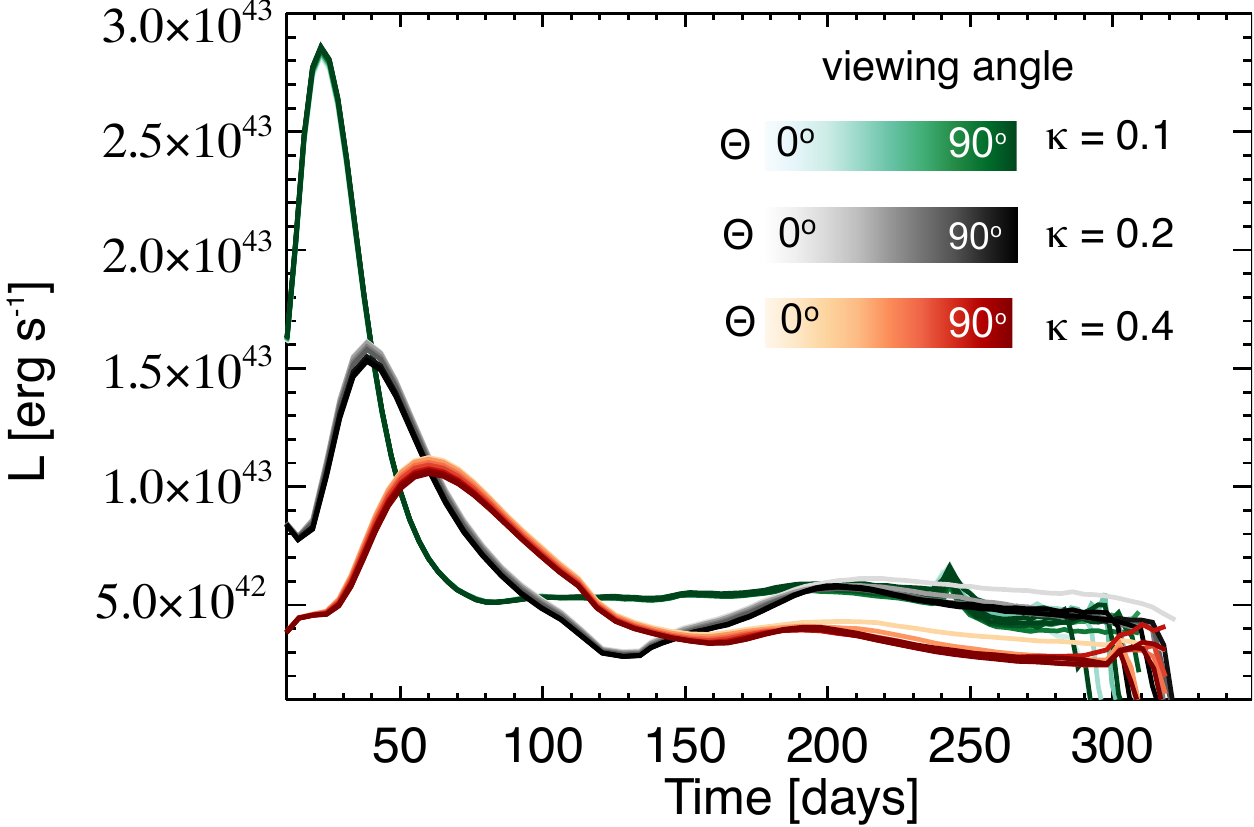} 
\caption{LCs for the 2D 512$^2$ AMR run with $\kappa =$ 0.1, 0.2, and 0.4.  Each graded color plot contains LCs from ten viewing angles: $\theta = 0^{\circ} - 90^{\circ}$ in $10^{\circ}$ increments.  The LCs peak at $1.1 - 2.8 \times 10^{43}$ erg s$^{-1}$ for $70-130$ days and then plateau at $3-5 \times 10^{42}$ erg s$^{-1}$ for  $150-200$ days. 
\label{fig:2d_lc}}
\end{center}
\end{figure}

We show images of density and radiation flux for the 2D PPI SN in Figure~\ref{fig:2d_den}. P2 and P3 again form dense shells because of radiative cooling soon after the simulation begins. They merge at 157 days and develop minor irregularities in structure with slight departures from spherical symmetry. We compare densities for our 2D RHD models with the 2D HD model from \citet{chen14a} at 50 days and 256 days in Figure~\ref{fig:comp2d}. The two models have similar resolution.  The spherical symmetry of both explosions begins to break down by 50 days. In the RHD model, radiative cooling in the postshock gas causes it to pile up into the thin dense shells in P3 and P2 at 2.05 $\times$ 10$^{15}$ cm and 2.45 $\times$ 10$^{15}$ cm, respectively. Cooling also decelerates the shell and prevents the formation of a reverse shock that creates the pressure inversion that drives RT instabilities. Instead, there are small ripples in the thin shell in the RHD model due to the preferential escape of radiation along lines of the sight with lower optical depths, but they are minor and more reminiscent of Vishniac overstabilities than instabilities \citep{v83}. At the same time, the collision in the HD model produces prominent RT instabilities that mix the ejecta behind the shock. By 256 days, the ripples in the shell in the RHD model have grown somewhat in amplitude but are still smaller than the mixing in the HD model. 

Figure~\ref{fig:2d_comp1} shows the shells at 270 days for different opacities and resolutions. At a given opacity, the 256$^2$ and 512$^2$ AMR runs indicate that the amplitude of the ripples and the thickness of shells decreases as the opacity increases.  Radiation can escape through ejecta at earlier times at lower opacities, so the deformation of the shells appears earlier and grows for longer times at a given resolution. \rev{The perturbations along the two axes are likely due to the carbuncle instability. Low-diffusivity hydrodynamics schemes like the high-order Godunov method in \CASTRO\ can be prone to such instabilities, which tend to arise where the shock is propagating along a coordinate axis \citep{quirk94}. Because they are zero-wavelength modes that arise at any resolution, no matter how high, their amplitude and width are not very dependent on resolution \citep[see, e.g., section 4.2 of][]{paper2}. Figure~\ref{fig:2d_comp1} shows this to be the case for the perturbations along the axis. However, the off-axis modes that also lead to the formation of hot spots in the shell are radiative instabilities because they do change with resolution. There, the higher grid resolution led to smaller perturbation amplitudes because the radiation diffusion time scales became smaller, so the colliding shells cooled sooner and pressure had less time to grow the perturbations. Furthermore, if we compare the dense shell in the 2D RHD runs with the fluid instabilities seen in Figure 5 of \citet{chen14a}, we see that the thin shell forms at the same location of the RT fingers in the hydro runs. This corroborates the point that the off-axis perturbations are fluid instabilities, not numerical effects. The dimple at $\sim 45^{\circ}$ is due to shadowing by a dense clump in the flow, it is also present in our 3D runs as shown in the next section.}

We show LCs for the 2D PPI SN for $\kappa =$ 0.1, 0.2, and 0.4 for the 512$^2$ AMR runs in Figure~\ref{fig:2d_lc}. The peak luminosities in 2D all exceed 10$^{43}$ erg s$^{-1}$, making the PPI SN brighter at early times than in 1D, and the second peak of the 1D LC is smoothed out in 2D. Furthermore, the luminosity of the LCs slightly change with viewing angles by up to 5\%, which can be explained by the ripples in the shell which emit anisotropic radiation flux.
 
\section{3D PPI SN Evolution}

% Figure 8

\begin{figure*}
\begin{center}
\includegraphics[width=\textwidth]{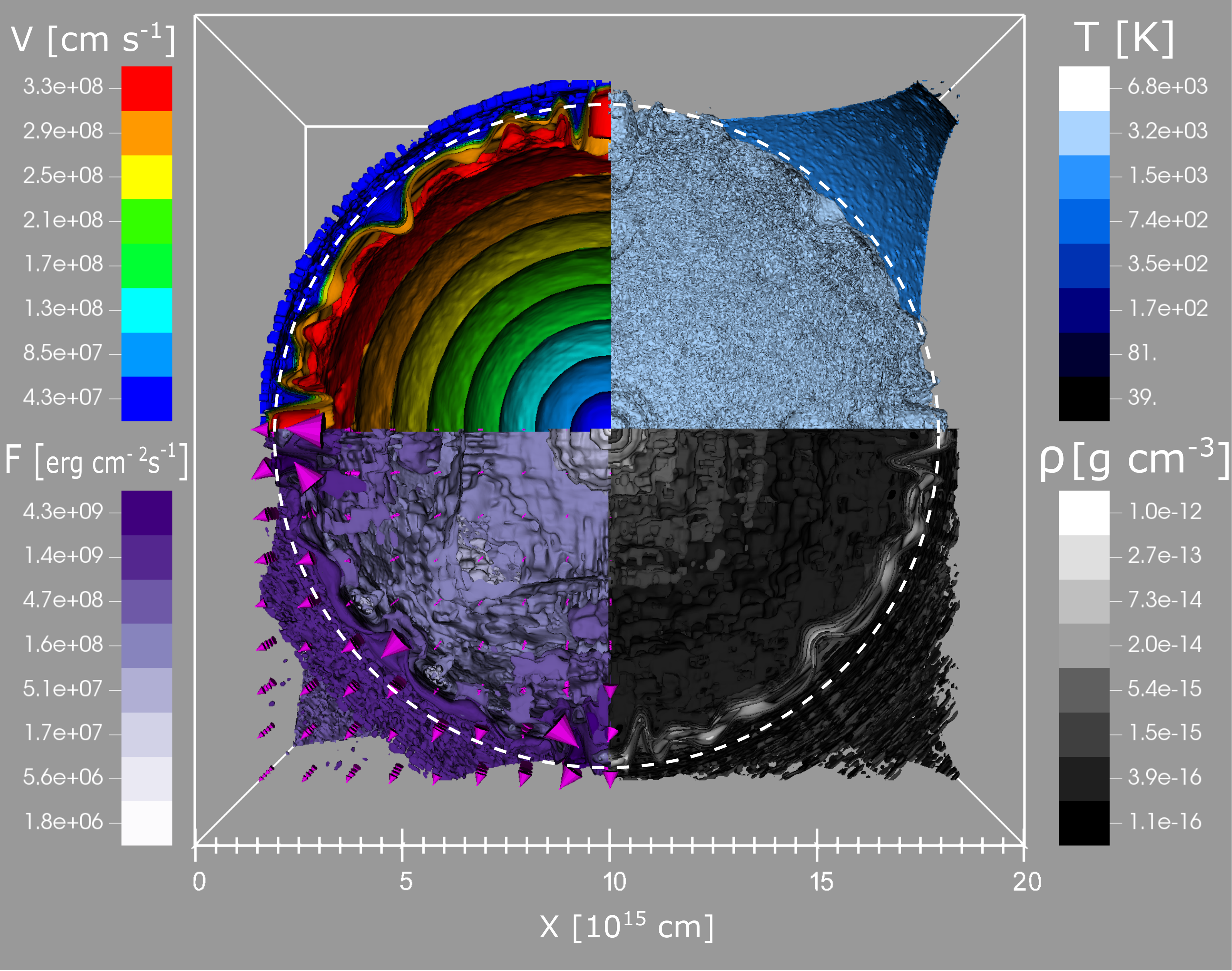} 
\caption{Snapshot of the 3D $\kappa =$ 0.2 run at 282 days. The four quadrants show radial velocity, gas temperature, radiation flux, and gas density. The pink arrows in the radiation flux show its direction. The white dashed circle marks where the shell begins to deviate from spherical symmetry.  
\label{fig:3d_flux}}
\end{center}
\end{figure*}

% Figure 9

\begin{figure*}
\begin{center}
\begin{tabular}{cc}
\includegraphics[width=.35\columnwidth]{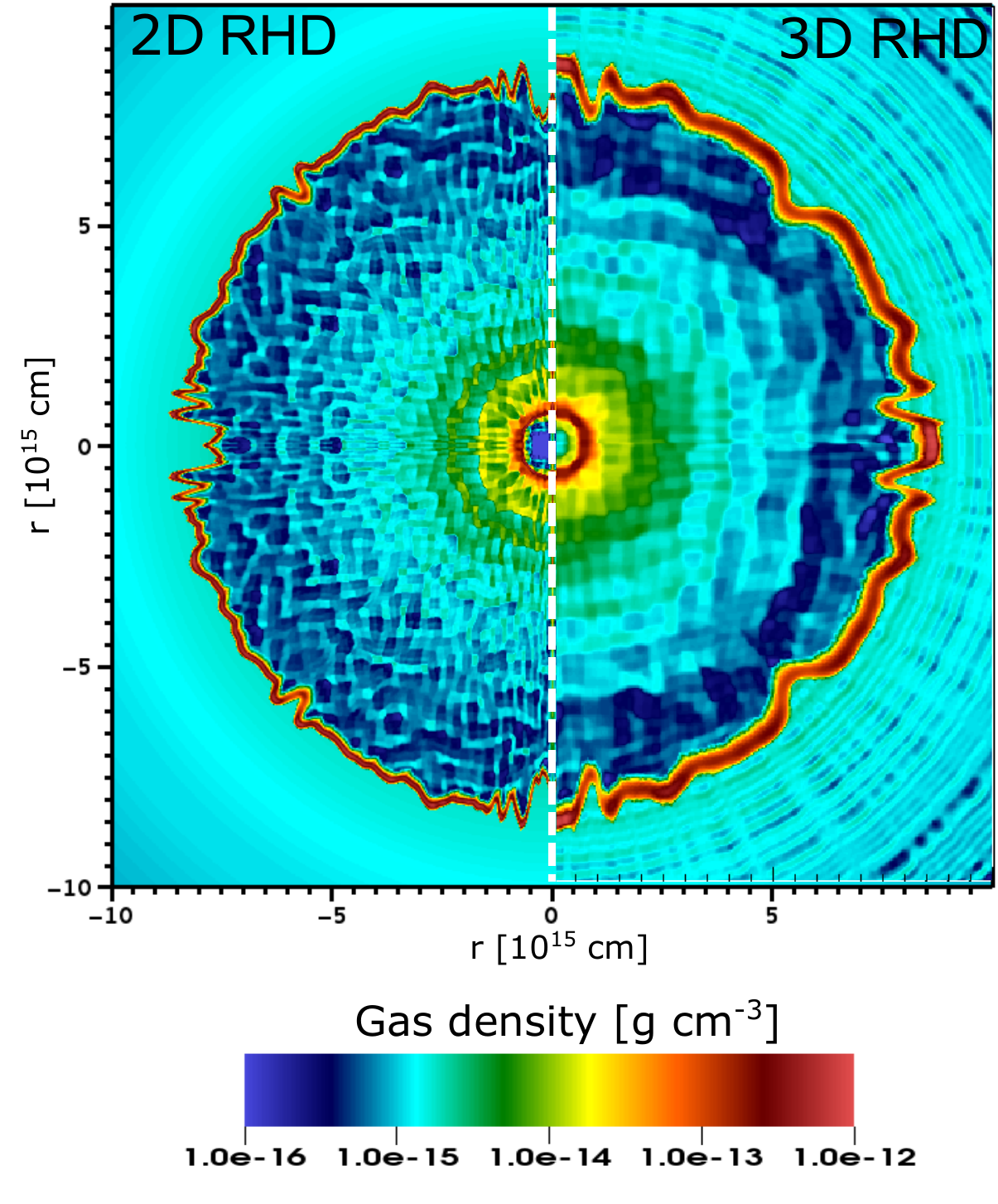}  &
\includegraphics[width=.55\columnwidth]{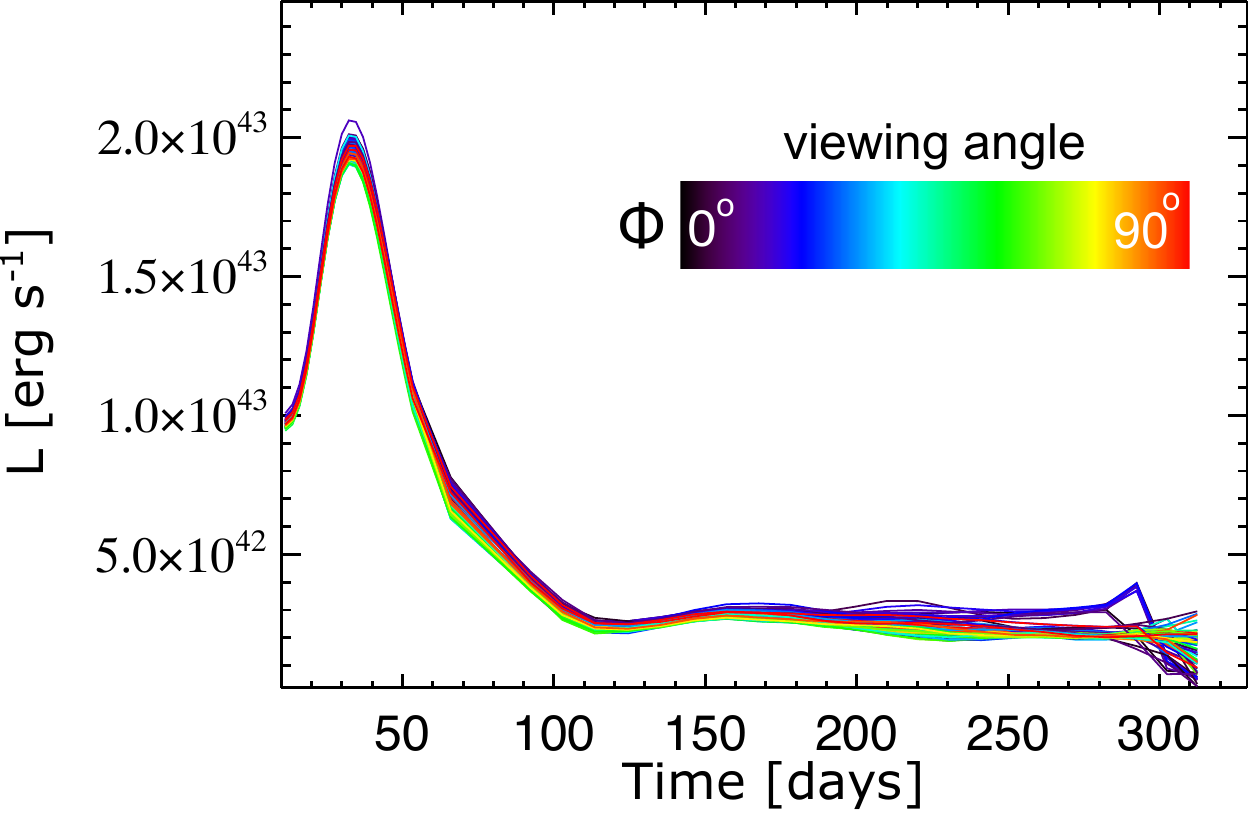} 
\end{tabular}
\caption{\emph{Left}: Density slices of the 2D and 3D runs for $\kappa =$ 0.2 at 250 days. Their respective resolutions are $4.88 \times 10^{12}$ and $9.76 \times 10^{12}$ cm.  The shell is at $r \sim 8 \times 10^{15}$ cm in both runs. More fine structure is visible in 2D than in 3D.  \emph{Right}: LCs for the 3D $\kappa =$ 0.2 model. The colors represent 100 LCs from ten random azimuthal angles, $\theta$ and ten polar angles, $\phi = 0^{\circ} - 90^{\circ}$ in $10^{\circ}$ increments.  These LCs sample luminosity variations due to viewing angle. The luminosity peaks at $1.9 - 2.1 \times 10^{43}$ erg s$^{-1}$ for $\sim$ 50 days and then plateaus ay $2-3 \times 10^{43}$ erg s$^{-1}$ for $\sim 200$ days. As in 2D, the 3D LCs do not have a second peak.  
\label{fig:2d_3d}}
\end{center}
\end{figure*}

Turbulent mixing in colliding shells can only be properly modeled in 3D. We show a snapshot of gas density, velocity, temperature, and radiation flux for the 3D $\kappa =$ 0.2 PPI SN at 282 days in Figure~\ref{fig:3d_flux}, when the shell is at $r \sim 8 \times 10^{15}$ cm.  Some turbulent flow is visible in the temperature and density images, and it can affect the later evolution of the core.  As in 2D, anisotropic radiation emitted from the shell causes only slight departures from spherical symmetry.  The LCs for the 3D explosion at different viewing angles are shown in the right panel of Figure~\ref{fig:2d_3d}. The luminosity peaks at $1.9-2.1 \times 10^{43}$ erg s$^{-1}$ for 110 days and then plateaus at $2-3 \times 10^{42}$ erg s$^{-1}$ for about 200 days.  Also as in 2D, the LCs in 3D lack the second peak seen in 1D. As the shell approaches the boundary of the box, the flux begins to vary with viewing angles. The LCs of the 2D and 3D simulations are similar, suggesting that 2D models capture their key features at a fraction of the computational cost of 3D runs.

\section{Discussion}

P3, P2, and some fraction of P1 collide at $10^{15}-10^{16}$ cm, where their relative kinetic energies are converted into thermal energy. Since this thermal energy is generated in the optically thin region, much of it radiates away and powers the luminosity of the SN. Our LCs indicate that the total radiation emitted by the collision is $1.22 - 1.64 \times 10^{50}$ erg. Since the total kinetic energy of P2+P3 is $\sim 6 \times 10^{50}$ erg, about $20 - 27\%$ of this energy is converted into radiation. This conversion efficiency can be estimated from the inelastic collision of shells in which momentum is conserved. If the two shells have masses and velocities $M_a$, $V_a$ and $M_b$, $V_b$, respectively, then the relative kinetic energy that is converted to heat and emitted as thermal radiation is
\begin{equation} \label{eq_col}
\begin{split}
E_r &= \frac{1}{2} M_a V^2_a + \frac{1}{2} M_b V^2_b - \frac{1}{2} (M_a+M_b) (\frac{M_a V_a+ M_b V_b}{M_a+M_b})^2  , \\
    & = \frac{1}{2}\frac{M_a M_b}{(M_a+M_b)}(V_a-V_b)^2. 
\end{split}
\end{equation}
Here, $M_a$ is the mass of P2+P3, $\sim$ 5.1 \Ms, $M_b$ is taken to be  20\% of the mass of P1, $\sim$ 5 \Ms\ and $V_a - V_b$ is $\sim 3\times 10^8$ cm s$^{-1}$. The total radiated energy is then $2.25 \times 10^{50}$ erg, which is consistent with our numerical results.

\begin{figure*}
\begin{center}
\begin{tabular}{cc}
\includegraphics[width=.45\columnwidth]{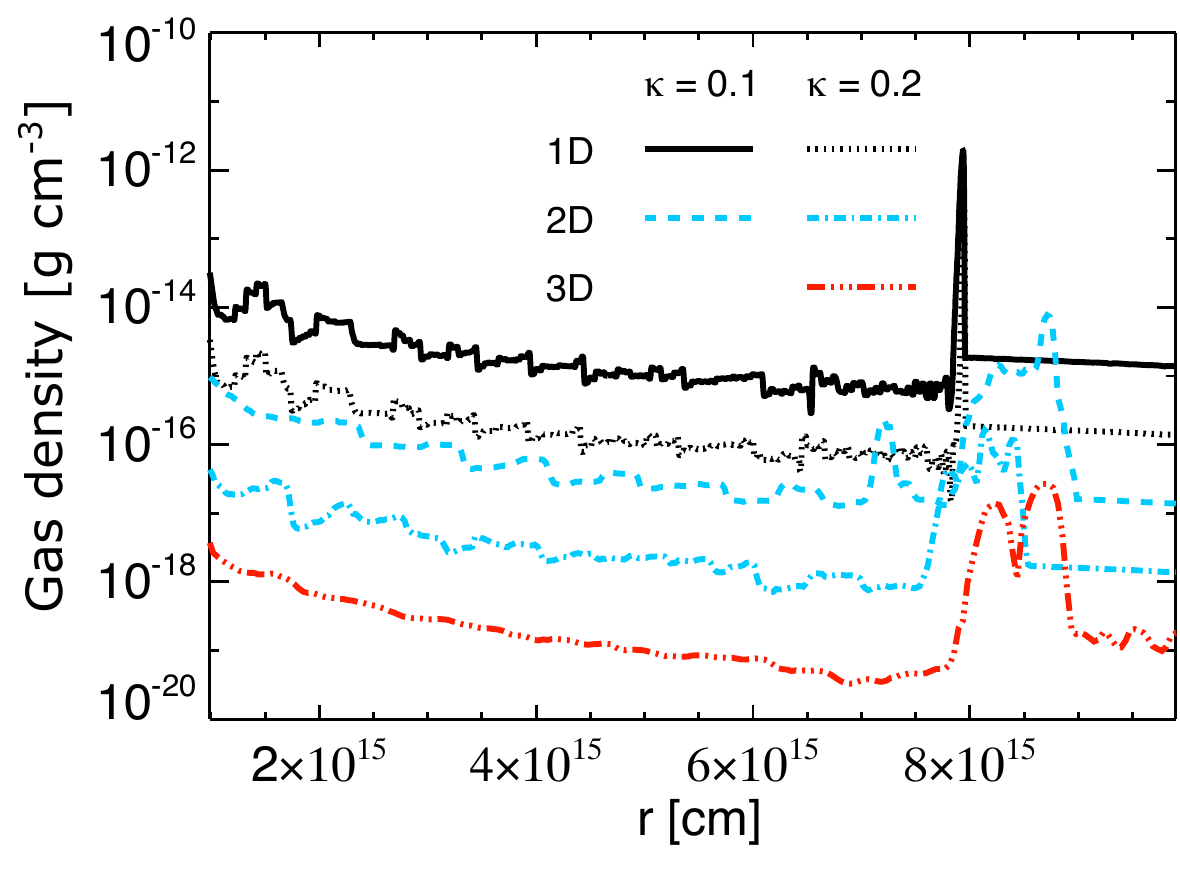}  &
\includegraphics[width=.45\columnwidth]{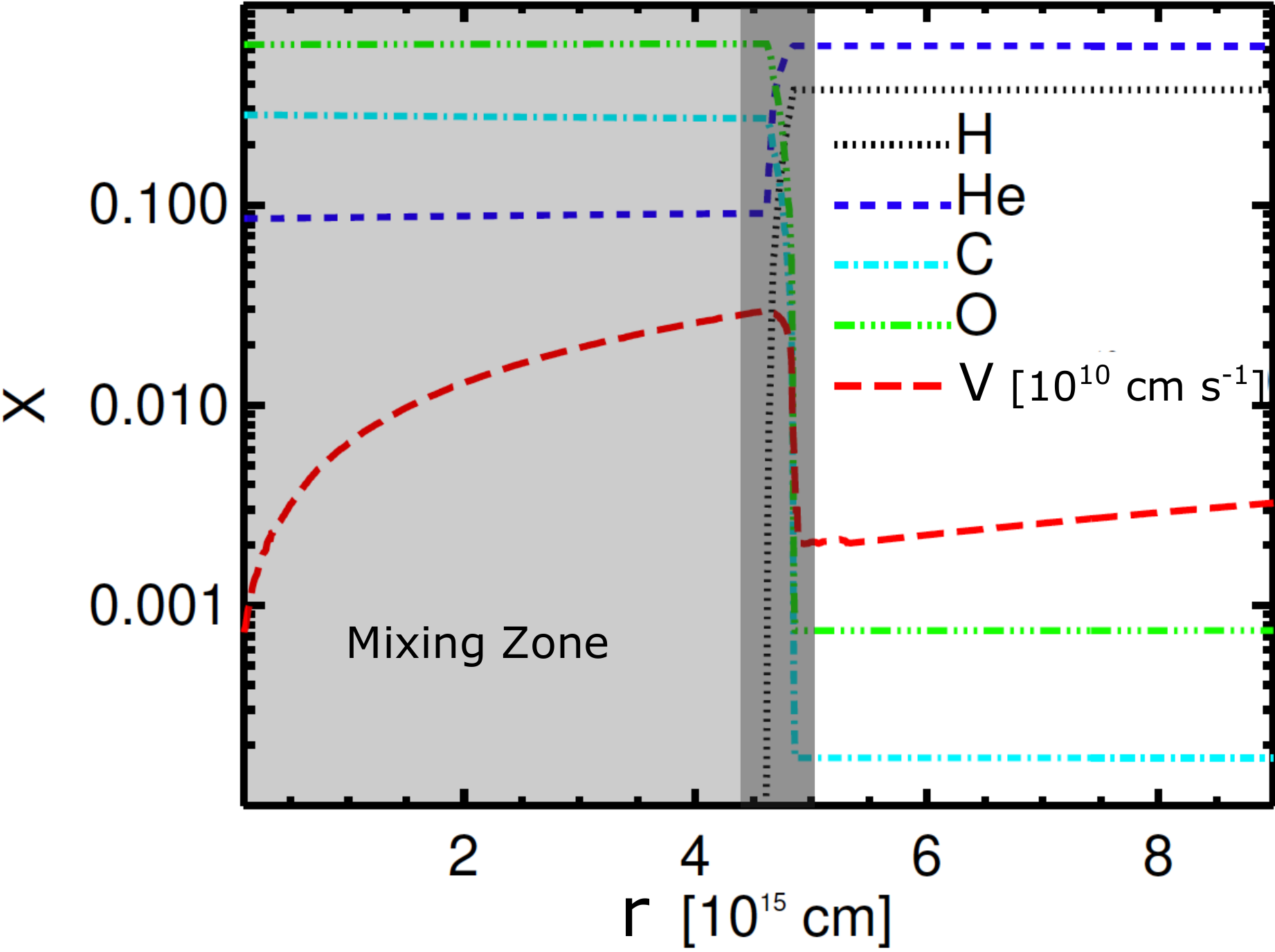} 
\end{tabular}
\caption{\emph{Left}: Angle-averaged density profiles for the 1D, 2D, and 3D $\kappa$ = 0.2 runs at 250 days, each offset by a factor of ten for clarity.  The density spike in 1D splits into multiple bumps in 2D and 3D.  \emph{Right}:  Angle-averaged mass fractions and velocities for the 2D $\kappa =$ 0.2 run at 135 days when the P2+P3 shell begins to form.  The dark gray band shows where \Cx\ and \Ox-rich ejecta in the shell mix with \Hy\ and \He\ in the wake of P1.  
\label{fig:angle_avgs}}
\end{center}
\end{figure*}

\rev{We compare density slices of the 2D and 3D RHD runs at 250 days in the left panel of Figure~\ref{fig:2d_3d}.  The shells are at the same radii in both runs and exhibit similar degrees of dimpling but the 3D shell is thicker and its interior has less turbulence, as shown by its smoother density structure. All else equal, small-scale turbulence is usually weaker in 3D than in 2D because the 2D turbulence is inversely cascaded. We compare 1D angle-averaged density profiles for the 2D and 3D runs to the 8192-zone 1D run at 250 days in the left panel of Figure~\ref{fig:angle_avgs}. The 1D spike forms regardless of the value of the opacity, with overdensities that exceed 1000 shortly after it appears. It fragments into multiple peaks with smaller overdensities of $\sim$ 100 in the 2D and 3D explosions. Although the shell densities vary with viewing angle in 2D and 3D, their overdensities are ten times smaller than those in the 1D run. These 1D angle-averaged plots exhibit broadened features similar to those in \citet{Duffell_2016}.}

We show angle-averaged profiles of elemental mass fractions and velocities for the 2D $\kappa =$ 0.2, 512$^2$ AMR run at 135 days in the right panel of Figure~\ref{fig:angle_avgs}. PPI SNe mostly eject elements lighter than \Si\ because the eruption is powered by explosive oxygen burning and it only ejects the outer part of the oxygen core. The high \Cx\ and \Ox\ mass fractions interior to the shell are primarily those in P3 at the time of ejection instead of being a product of mixing because radiative cooling suppresses fluid instabilities. Minor mixing of \Cx\ and \Ox\ with \Hy\ and \He\ occurs in the dark gray region of the shell shown in the right panel of Figure~\ref{fig:angle_avgs}, and these elements would be visible to an external observer. Since the shell contains \Cx\ and \Ox-rich ejecta, PPI SNe are expected to exhibit prominent \Cx\ and \Ox\ lines but only weak \Si\ or \Fe\ lines that are due to the initial metallicity of the progenitor star.

\begin{figure*}
\begin{center}
\includegraphics[width=\textwidth]{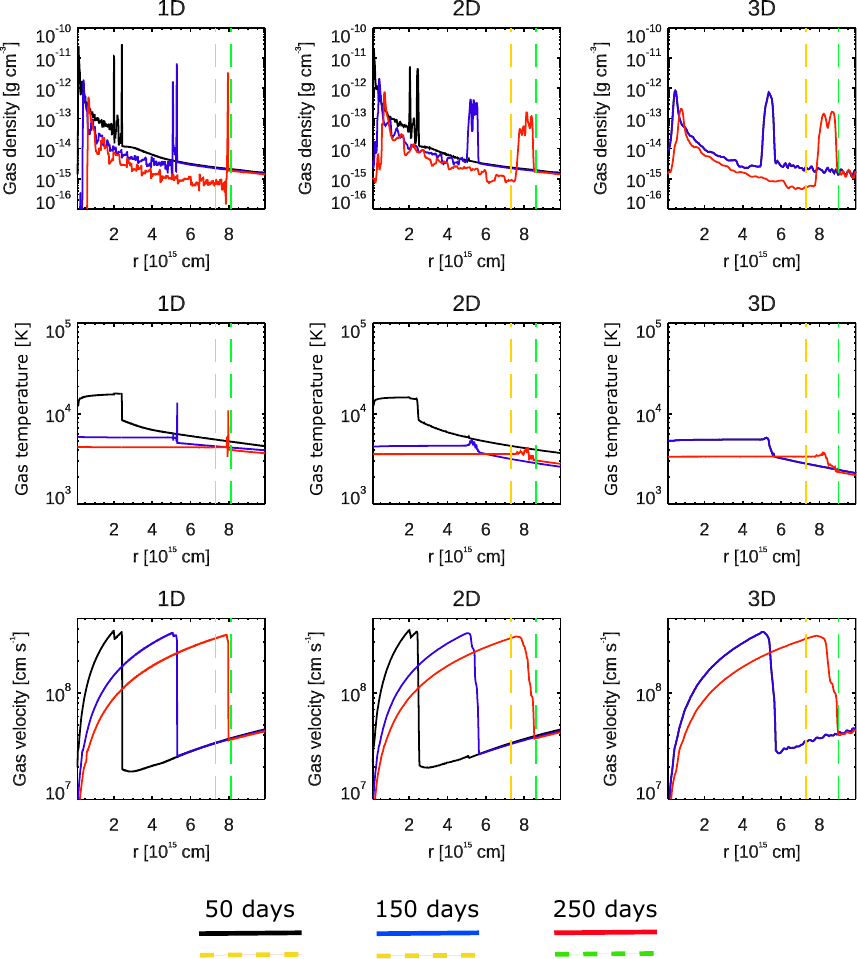} 
\caption{ \rev{Comparison of 1D angle-averaged profiles of gas density, temperature, and velocity for the 2D and 3D runs with the 1D run at 50, 150, and 250 days, respectively, for $\kappa =$ 0.2 at the highest resolutions. The vertical dashed yellow and green lines indicate the locations of the photosphere at 50, 150, and 250 days, respectively. There is only one yellow line because the photosphere has not moved from 50 - 150 days.}  
\label{fig:ang_all}}
\end{center}
\end{figure*}

\rev{We compare 1D angle-averaged profiles of density, temperature, and velocity for the 2D and 3D runs to those of our 1D runs at 50, 150, and 250 days in Figure~\ref{fig:ang_all} for $\kappa =$ 0.2 and the highest resolutions (plots at 50 days for the 3D run do not appear because data was not output at that time in the run).  The 1D and 2D temperature profiles show that radiation is still trapped behind the shock in the ejecta at 50 days but has mostly leaked out by 150 days, when temperatures behind the shock have become flat. The reverse shock is clearly visible at 50 days in the velocity profiles but only slightly at 150 days in the 1D run and not at all in the 2D run, in which radiative instabilities have smeared it out. This sequence shows how radiative cooling dissipates the energy of the reverse shock at early times, which suppresses the formation of RT instabilities.}  

\rev{The location of the photosphere in all three runs is shown by the vertical yellow and green dashed lines. It is at the same location at 50 and 100 days in all three runs and is well beyond the shock at both times. This is due to the fact that we assume the same opacity throughout the gas for simplicity, whether or not it is really ionized. The luminosity at the peak of the LCs at 10 - 50 days thus had to diffuse out to larger radii before becoming visible, typically on timescales of a day or two.  The differences in the LCs between the runs at early times are therefore not due to differences in photospheric radii. Our 2D and 3D simulations produce higher peak luminosities than 1D simulations because radiative cooling leads to the formation of hot spots on the shell due to fluid instabilities.  At late times, the growth of the instabilities becomes saturated and the hot spots cool via radiation diffusion so there is better agreement between the LCs. The hot spots due to corrugations in the shell also introduce the variations in luminosity with viewing angle visible in Figures~\ref{fig:2d_lc} and \ref{fig:2d_3d}, but they are at most only $\sim$ 5\%. This effect is minor in comparison to the overall brightening of the collision they produce.}

\begin{figure*}
\begin{center}
\begin{tabular}{c} 
\includegraphics[width=1.\textwidth]{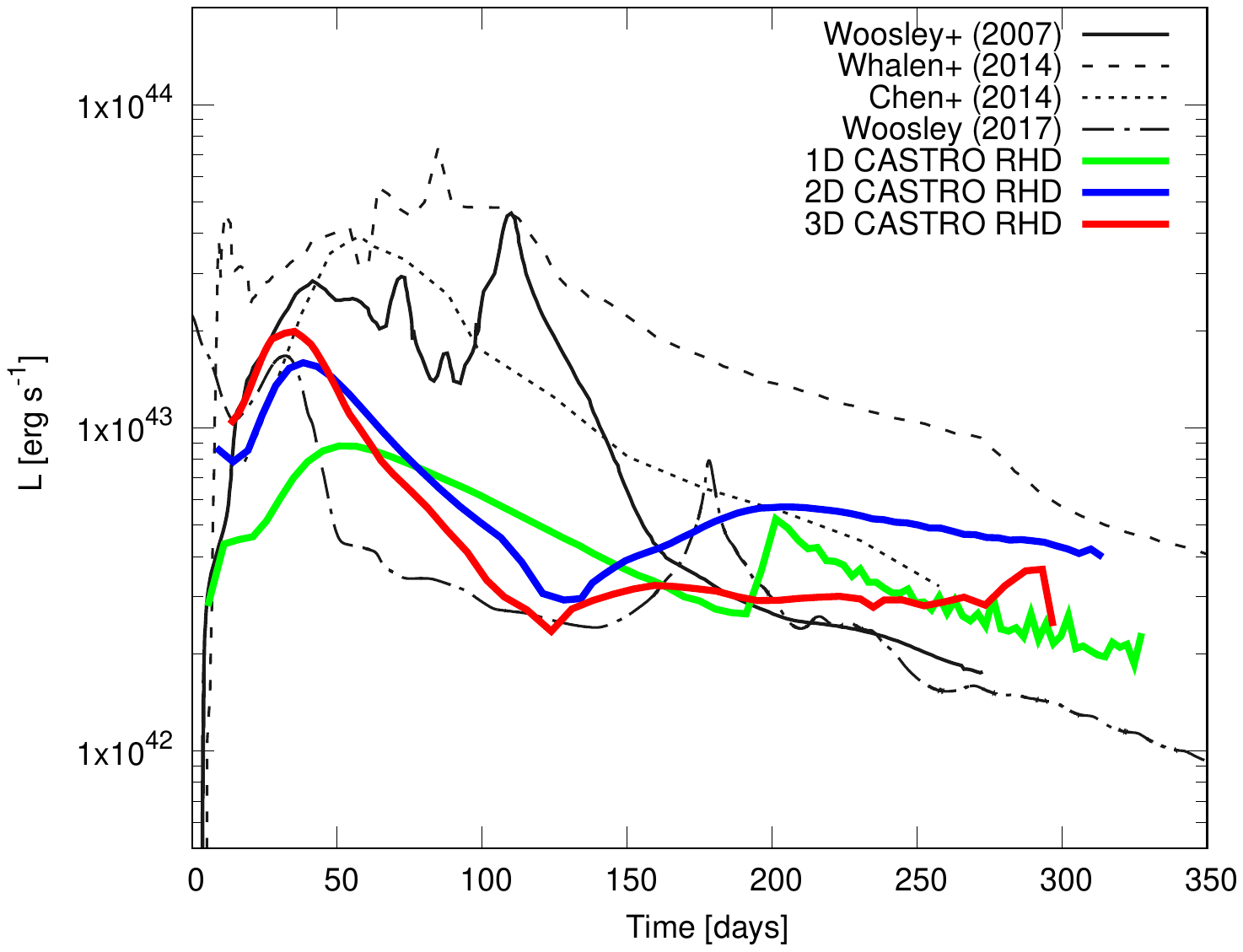} 
\end{tabular}
\caption{\rev{Comparison of our 1D, 2D and 3D LCs $\kappa =$ 0.2 \CASTRO\ runs with the highest resolutions to those of previous studies. The absence of secondary peaks in our 2D and 3D simulations highlight that multidimensional RHD simulations are required for realistic PPI SN LCs.} 
\label{fig:all_lcs}}
\end{center}
\end{figure*}

\rev{We compare our LCs to those of previous studies in Figure~\ref{fig:all_lcs}. \citet{wbh07,wet13a,w17} produced 1D LCs with multiple peaks with $\sim$ $2.0 - 6.9 \times 10^{43}$ erg s$^{-1}$ that are broadly consistent with those in our LCs. In particular, our 1D LC is a good match to that of the T110B model in \citet{w17}. However, the LCs of \citet{wbh07,wet13a,w17} exhibit slowly-decaying luminosities at late times while ours suggest that they are more like a plateau. This discrepancy is likely due to opacities, as our simple constant opacities neglect the detailed ionization state of the colliding shells. The absence of the second peak in our 2D and 3D models nevertheless highlight that multidimensional RHD simulations are required to compute realistic PPI SN LCs.}
  
% Figure 10

% Figure 11

% Figure 12

\section{Conclusion}

\rev{Our simulations clearly show that multidimensional simulations with radiation transport are required to capture the true evolution of PPI SNe their LCs. They show that the dense shell  from which most of the luminosity of the collision originates in 1D models fractures into dense clumps with hot spots that can enhance peak luminosities by factors of 2 - 3 at early times, and that mixing in the ejecta eliminates the second peak found in most 1D LCs. Radiation transport dampens the violent RT instabilities that occur in 2D models without RHD, producing milder radiative overstabilities without much mixing.  LCs from our 3D high-resolution runs peak at $1.9 - 2.1 \times 10^{43}$ erg s$^{-1}$ for $\sim$ 50 days and then plateau at $\sim$ $2 - 3 \times 10^{43}$ erg s$^{-1}$ for $\sim$ 200 days, depending on viewing angle. The first peak is powered by collisional heating between P2 and P3 and the plateau is powered by the merged P2+P3 shell plowing up the tail of P1. Although there are some morphological differences in the ejecta in 2D and 3D, most of the salient features of the collision and its LC can be captured in 2D RHD simulations at a fraction of the cost of 3D runs. Our 1D, 2D, and 3D \CASTRO\ models suggest that multidimensional simulations with radiation transport are required to understand the evolution of all shell-collision SNe such as Type IIne, not just PPI SNe, because of similarities in their dynamics.}

\rev{Although peak luminosities are higher in our multidimensional runs, the total energy emitted by the collision is essentially the same as in 1D: $1.22 - 1.64 \times 10^{50}$ erg, or $\sim$ 20 - 27\% of the total kinetic energy of P2 and P3. Because radiative cooling due to photon transport weakens the reverse shock that forms during the collision and suppresses the growth of RT instabilities, mixing is much less violent in our models than in previous HD-only simulations. But even though not much mixing occurs, the collision of P3 and P2 will yield \Cx\ and \Ox\ emission lines because P3 carries these species into the merged shell. This together with the absence of \Si\ and\Fe\ lines could be used to identify a transient as a PPI SN for follow-up observations.}

\rev{The goal of our paper was to model how the evolution of PPI SNe changes with radiation hydrodynamics in 2D and 3D and how those changes would affect its LC for a reasonable set of physics, not produce sophisticated LCs for observational proposals. However, our models could be improved with more realistic opacities, such as electron fractions calculated with multi-species Saha equations or comprehensive opacity tables \citep{opal1,opal3}. Higher-order RHD schemes in which the angular distribution of the radiation field is calculated at each point in space, such as implicit Monte Carlo \citep[IMC;][]{kasen06} or variable Eddington tensor formalism \citep[VETF;][]{vet}, are better suited to the dense clumps in the disrupted shell because they capture shadowing better than FLD, wherein radiation can unphysically flow around obstacles. In the future we will build new RHD models of PPI SNe with progenitor stars from \citet{w17} in order to better predict their observational signatures.}

The detection and proper characterization of PPI SNe will open new windows on massive star formation in the local and primordial universe.  The advent of new SN factories such as the {\it Vera Rubin Telescope} and wide-field near-infrared missions such as the {\it Nancy Grace Roman Space Telescope} and {\em Euclid} will reveal more of these events and enlarge our understanding of PPI SNe and the deaths of very massive stars in the coming decade.  

\section{Acknowledgement}

We thank the anonymous referee, whose critique improved the quality of this work.  This research is supported by the Ministry of Science and Technology, Taiwan under grant no. MOST 110-2112-M-001-068-MY3 and the Academia Sinica, Taiwan under a career development award under grant no. AS-CDA-111-M04.  Our computing resources were supported by the National Energy Research Scientific Computing Center (NERSC), a U.S. Department of Energy Office of Science User Facility operated under Contract No. DE-AC02-05CH11231,  the Center for Computational Astrophysics (CfCA) at National Astronomical Observatory of Japan (NAOJ),  and the TIARA Cluster at the Academia Sinica Institute of Astronomy and Astrophysics (ASIAA).

%\bibliographystyle{aasjournal}
%\bibliography{refs}

\end{document}